\documentclass[epj-spec]{svjour}
\usepackage{graphics,graphicx}
\usepackage{subfigure}
\usepackage{amsfonts}
\usepackage{amsmath}
\usepackage{amssymb}

\newcommand {\be} {\begin{equation}} 
\newcommand {\ee} {\end{equation}} 
\newcommand {\Be}{\begin{eqnarray*}}
\newcommand {\Ee} {\end{eqnarray*}}
\newcommand {\bey} {\begin{eqnarray}} 
\newcommand {\eey} {\end{eqnarray}} 

\begin{document}

\title{Energy diffusion in hard-point systems}

\author{Luca Delfini \inst{1} \and Sergey Denisov \inst{2} \and 
Stefano Lepri \inst{1} \and Roberto Livi \inst{3} 
\thanks{Also at Istituto Nazionale di Fisica Nucleare 
and Istituto Nazionale di Fisica della Materia, Firenze.} \and
Pradeep K. Mohanty \inst{4}
\and Antonio Politi \inst{1}
}                     
\institute{Istituto dei Sistemi Complessi, Consiglio Nazionale
delle Ricerche, via Madonna del Piano 10, I-50019 Sesto Fiorentino, Italy.
\and  Department of Physics, University of Augsburg
D-86135 Augsburg Germany
\and  Dipartimento di Fisica, via G. Sansone 1 I-50019, Sesto
Fiorentino, Italy
\and Saha Institute of Nuclear Physics, Kolkata, India
}

\date{Received: \today}

\abstract{ 
We investigate the diffusive properties of energy fluctuations in a
one-dimensional diatomic chain of hard-point particles interacting through a
square--well potential. The evolution of initially localized infinitesimal and
finite perturbations is numerically investigated for different density values.
All cases belong to the same universality class which can be also interpreted
as a Levy walk of the energy with scaling exponent $\gamma=3/5$. The
zero-pressure limit is nevertheless exceptional in that normal diffusion
is found in tangent space and yet anomalous diffusion with a different rate
for perturbations of finite amplitude. The different behaviour of the two
classes of perturbations is traced back to the ``stable chaos" type of dynamics
exhibited by this model. Finally, the effect of an additional internal degree of
freedom is investigated, finding that it does not modify the overall scenario.}

\maketitle  

\section{Introduction}

Diffusion in one--dimensional systems of interacting particle may display
peculiar features. For instance, consider an assembly of impenetrable spheres
confined within a narrow channel. If the mutual passage of particles is
forbidden, the motion of the spheres is necessarily correlated, even at long
times, because the displacement of a given particle over a long distance
necessitates a rearrangement of many other particles in the same direction. 
This is an observed effect, for example, in single-filing systems where 
Fick's law is violated ~\cite{singlef}. 

The distinguished signature of those effects is in the long-time behavior of
the associated correlation functions \cite{PR75}. As it is known, the latter
may display long--time tails leading to ill--defined transport coefficients or,
more generally, to the breakdown of customary hydrodynamics. Indeed, power--law
decay of correlations is expected to be a generic feature of one--dimensional
systems in the presence of conservation laws ~\cite{NR02}. One important
consequence of such long--ranged correlations is that physical properties may
significantly depend on the system size and that the thermodynamic and
infinite--time limits may not commute. For instance, the tagged--particle
diffusion coefficient in $D=1$, that is finite for the infinite system is
found instead to vanish for a finite one (see \cite{pal} and references
therein). 

Another important example is the divergence of the thermal conductivity
coefficient $\kappa$  observed in chains of anharmonic
oscillators~\cite{LLP97}. It is now assessed \cite{LLP03} that in 1D 
$\kappa$ diverges with the system's length $L$
\be
\kappa \sim L^{\alpha}
\label{kappa}
\ee
A considerable number of papers have reported anomalies of this type and we
refer the reader to recent review articles \cite{LLP03,chaos}  for an account of
the literature. Related manifestation of such anomalies are the divergence of
viscosity in  $1D$ lattice gases \cite{DLQ89} and the anomalous scaling of
Rayleigh  and Brillouin peak widths in the hydrodynamic limit detected for the
Lennard--Jones fluid \cite{Bishop2,sandri}.

For what concerns the actual values of $\alpha$ there are still some
controversies.  The renormalization--group arguments of Ref.~\cite{NR02} give
$\alpha =1/3$. More recently, a critical revision of the self--consistent 
Mode-Coupling Theory approach \cite{schirm,L98} has confirmed this prediction
whenever the lowest nonlinear term in the theory is \textit{cubic} in the
relevant correlators \cite{DLLP06}. On the other hand, when the first
nonlinearity is \textit{quartic}, the value of the exponent turns out to be
$\alpha = 1/2$ \cite{canada,futuro}. Physically, this latter condition can be
achieved when the interaction potential is symmetric with respect to the
equilibrium position, and corresponds, more generally, to the case of
zero pressure. 

Many efforts have been made to verify numerically the theoretical predictions.
Simulations of an anharmonic chain with cubic plus quartic non-linearity yield
$\alpha$ = 1/3 \cite{DLLP06} and the same result has been obtained for a 1D 
gas of elastically colliding particles \cite{denis} with a great accuracy. 
On the other hand, the
same analysis carried out for chains with quartic nonlinearities (e.g. 
the FPU-$\beta$ model), although exhibits a
different scaling behaviour is not able to convincingly confirm the above
mentioned theoretical predictions \cite{LLP03b,DLLP06}.

Computer simulations of simple toy models is an invaluable way of attacking
those problems. In particular, one would like to understand the conditions under
which anomalies occur and to classify possibly universal features.  In this
sense, the Hard--Point Gas (HPG) \cite{C86}, made of point particles that
interact through elastic collisions, proved very helpful in clarifying some
aspects of heat conductivity in one-dimensional systems
\cite{H99,D01,grass,casati,hurtado}. A limitation of the HPG model is that one
cannot tune the pressure and, in particular it is not possible to test whether
the zero pressure limit indeed corresponds to a different universality class. In
order to overcome such limitations, in this paper we consider an extension of
the HPG model in which the distance between neighbouring  particles is also
limited from above by a square well potential. For reference, we will denote
this model as the \textit{hard--point chain model}.  Savin and collaborators
\cite{savin} have already shown that this systems displays anomalous properties.
What is  more important for our purposes is that it an equilibrium state with
vanishing pressure and we can study how  anomalous features change approaching
this limit. It should be also recalled that in Ref.~\cite{savin} only
the zero-pressure case was considered.

We indeed confirm the zero-pressure case belongs to a different universality
class, although it is again difficult to accurately determine the corresponding
scaling behaviour. Moreover, by studying both infinitesimal and finite
perturbations we convincingly find different properties which we interpret as a
signature of ``stable chaos", i.e. of an irregular behaviour induced by
discontinuities in phase space, rather than from the exponential amplification
of infinitesimal perturbations \cite{stcao}. We must indeed  recall that both
the HPG and the hard-point chain are not chaotic in the usual sense,  as the
maximum Lyapunov exponent strictly vanishes.

In section 2, the model and its physical parameters are introduced. In section
3, the diffusion properties of both infinitesimal and finite amplitude
perturbations are investigated for different values of the specific volume (i.e.
upon indirectly varying the pressure). The different velocities, that can be
defined dynamically and thermodynamically, are reviewed in Sec. 4. The
properties of the energy current are briefly discussed in Sec. 5 to  confirm the
indirect observation of an anomalous behavior at zero pressure. In Sec. 6 the
effect of an additional internal degree of freedom is analysed, while the last
section is devoted to summarizing the results and the open problems.

\section{The hard--point chain model}

\begin{figure}[!ht]
\begin{center}
\includegraphics[clip,width=8cm]{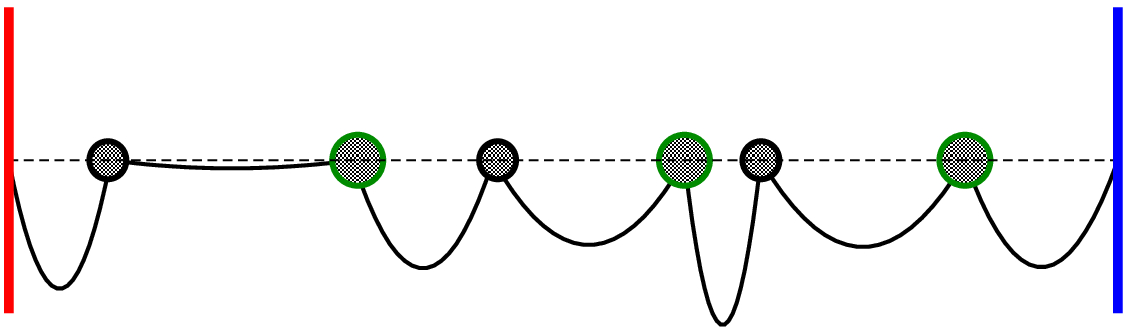}
\caption{Sketch of a diatomic hard-point chain. Different radii are suggestive
of different masses of particles that are assumed to be point-like.}
\label{hard_chain}
\end{center}
\end{figure}

The model consists of a set of $N$ point--like particles of masses $m_i$ and 
positions $x_i$ ordered along a line. Interactions are restricted only to
nearest-neighbour pairs with a square--well potential in the relative distances
\begin{displaymath}
 U(x_{i+1}-x_i)= \left \{ \begin{array}{ll}
     0 & \quad \mbox{$0\;<x_{i+1}-x_i\;<\;a$}\\
     \infty & \quad \mbox{otherwise}
     \end{array} \right.
\label{u}     
\end{displaymath}
The two infinite barriers at $x_{i+1}-x_i=0$ and $x_{i+1}-x_i=a$ correspond to
elastic ``collisions", the latter ones occurring at a finite distance as if
the particles were linked by an inextensible and massless string of fixed
length $a$ (see Fig.~\ref{hard_chain} for a pictorial image). The
string has no effect on the motion unless it reaches its maximal length, when
it excerts a restoring force that tends to rebounce the particles one 
against the other. Clearly, the potential (\ref{u}) introduces the physical 
distance $a$ as a parameter of the model.

Both types of collisions are described by the same updating rule for the
velocities $u_i=\dot x_i$, 
\begin{equation}
u_i'=\frac{m_i-m_{i+1}}{m_i+m_{i+1}} u_i+ \frac{2 m_{i+1}}{m_i+m_{i+1}}
u_{i+1} \quad , \quad
u_{i+1}'= \frac{2 m_{i}}{m_i+m_{i+1}}
u_i-\frac{m_i-m_{i+1}}{m_i+m_{i+1}} u_{i+1}, 
\label{coll}
\end{equation}

We consider the case of alternating masses, whose ratio is denoted by $r$,
i.e. $m_{i}=m$ for even $i$ and $m_{i}=rm$~, otherwise. This choice is 
dictated by
the need to break the integrability of the model that holds for $r=1$.
The particles are confined inside a ``box" of length $L$. We consider both
periodic boundary conditions 
\begin{equation}
x_{n+N} \;=\; x_{n} \,+\, L  \quad .
\label{pbc}
\end{equation}
and fixed boundary conditions,
\begin{equation}
x_1\;=\;0 \qquad x_N \;=\; L
\end{equation}
Accordingly, the specific volume $\ell=L/N$ is a state variable to be
considered together with the specific energy (energy per particle) 
$e= \sum_i m_{i}u^2_{i}/2N $.

\begin{figure}[!ht]
\begin{center}
\includegraphics[clip,width=8cm]{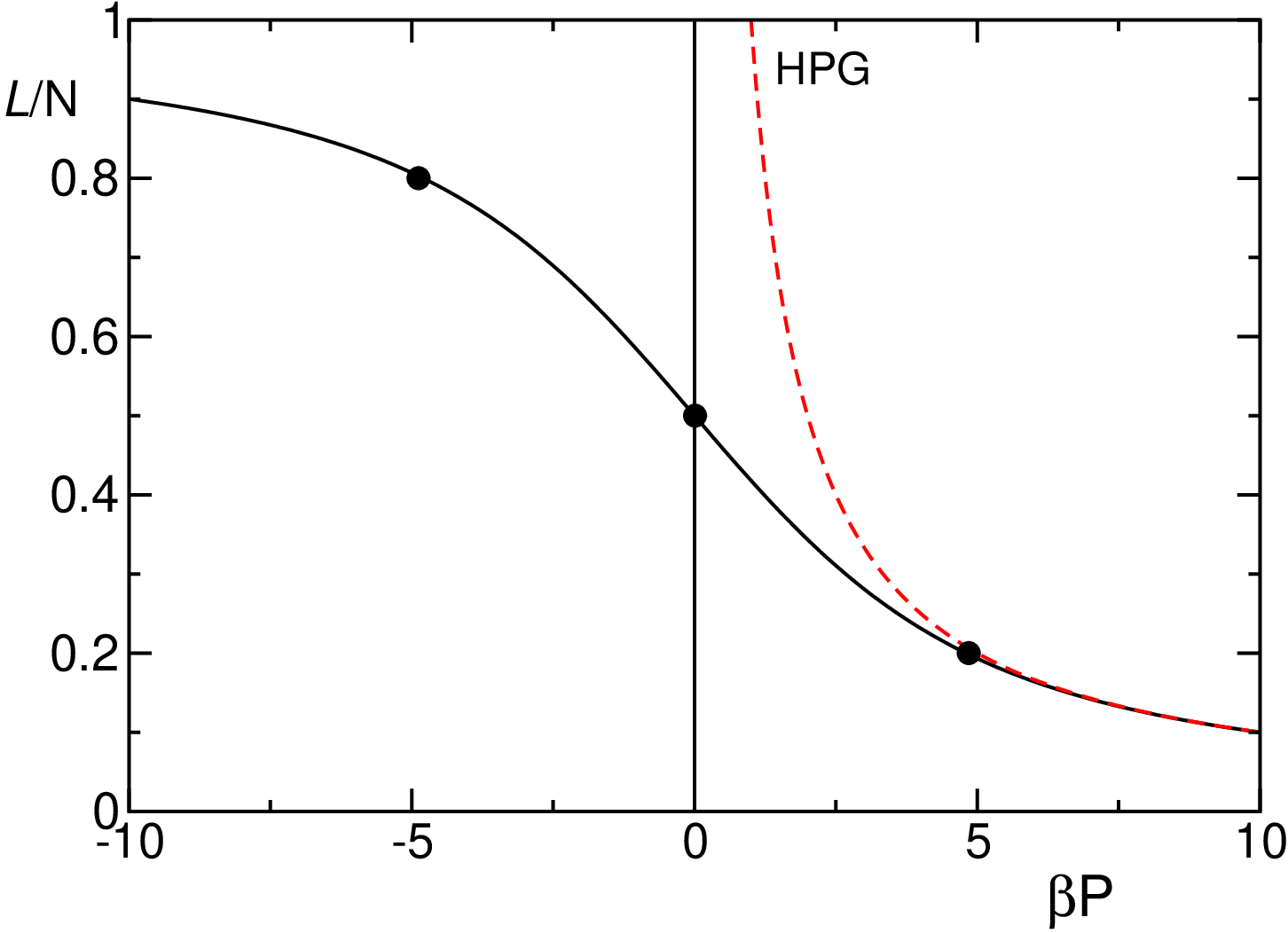}
\caption{Equation of state of the hard-point chain: the solid line 
is the analytic expression of eq.\ref{eqst}, points are obtained from 
numerical simulations with $e=1$.}
\label{eqstato}
\end{center}
\end{figure}

As it is well known, the thermodynamics of models like ours can be solved
exactly (see e.g. \cite{Bis}). In particular, the partition function and the
equation of state in the $P-T$ ensemble is 
\begin{equation}
Z = \prod_{j=1}^{N} \int_{-\infty}^{\infty} dp_{j} \int_{0}^{a} dx_{1} 
\int_{x_{1}}^{x_{1}+a}dx_{2}...\int_{x_{N-1}}^{x_{N-1}+a}dx_{N} \, 
\exp[-\beta \sum_{i}\frac{p_{i}}{2m_i} - Px_{N}]
\label{eq:state}
\end{equation}
where $p_i=m_i u_i$ are the conjugate momenta. Introducing the new variables
$\xi_i=x_{i+1}-x_i$ we find, after some manipulations 
\be
Z = \lambda^{-N} \left [ \int_{0}^{a}d\xi \exp(-\beta P \xi) \right ]^{N}
=\left [\frac{1}{\lambda \beta P} (1-\exp(-\beta P a))\right ]^{N}
\ee
where $\lambda$ is the De Broglie wavelength. The equation of state is
given by
\be
L = - \frac{ \partial \ln Z}{\partial (\beta P)} = 
N \left [ \frac{1}{\beta P}- \frac{a}{\exp(\beta Pa)-1} \right ]
\label{eqst} 
\ee 
This equation is shown in Fig.~(\ref{eqstato}) for $a = 1$.  For convenience we
report the specific volume $\ell$ as a function of $\beta P$. Note that for
large $a$ values, the equation of state is the same for an HPG (red line)
i.e. the one of an ideal gas in 1D. As a check of our codes (see below) 
we also computed some state parameters by fixing $\ell$ (namely the box length)
and evaluating the pressure as the average exchanged momentum between 
neighbouring particles.

Before discussing the results, let us briefly describe the simulation
procedure. Models
like the one we are dealing with are particularly suitable for numerical
computation as they do not require integration of nonlinear differential
equations. Indeed, the dynamics amounts simply to evaluating successive
collision times and updating the velocities according to Eqs.~(\ref{coll}). The
only errors are those due to machine round-off. Moreover, the simulation can be
made very efficient resorting to fast updating algorithms. The fact that the
$x_i$ contribute only indirectly to the evolution, by determining the collision
times together with the conservation of the particle ordering along the chain,
allows simulating the dynamics with an event driven algorithm that exploits the
heap structure of the future collision times \cite{grass}.

Without loss of generality, we can fix $m=1$ and $a=1$. Most runs have been
performed for a mass ratio $r=2$ and for an energy per particle $e=1$, which
corresponds to $\beta^{-1}=k_{B}T = 2$. Moreover, we have chosen to work
with two representative values of the specific volume, namely $\ell=0.2$ and
$\ell = 0.5$. In the former case, the particles hardly feel the barrier at
relative distance $1$, mainly interact only through standard collision. In this
regime, the  hard--point chain behaves like a gas and is indeed practically
equivalent to the HPG. On the contrary, for $\ell=0.5$, the two types of
collisions are equally probable, so that the pressure is zero.

\section{Diffusion of perturbations}
 
In the spirit of linear response theory \cite{forster}, transport coefficients 
can be determined by looking at the way a statistical system reacts to small
perturbations (external forces). For heat conductivity, one can
directly follow the evolution of an initially localized energy perturbation
$\triangle e(x,t)$ \cite{helfand}. If the perturbation is weak enough 
for the system to be locally in thermodynamic equilibrium, a reasonable
measure of the local temperature field $\triangle T(x,t)$ is $\triangle e(x,t) =
C_{P} \triangle T(x,t)$, where $C_{P}$ is the heat capacity per unit length at
constant pressure. Thus, by measuring the spreading of the perturbation field,
one can estimate how heat propagates through the system \cite{helfand}. A
standard diffusive process is the indication of a normal heat conductivity.

In general, quantities like $\triangle e(x,t)$ are determined by all particles
contained in a small box around the spatial location $x$. If, like in our 1D
model, particles cannot cross each other, it is more practical to identify $x$
as the average position $i\ell$ of the $i$-th  particle. To some extent, this
amounts to interpreting the model as a lattice system. Of course, nothing
changes in the underlying physics, provided that one consistently adopts the
same interpretation throughout. Moreover, since we are mainly interested in the 
long--time and large--distances behaviours this should not be very relevant,
anyhow.

With the above remark in mind, we consider a system at equilibrium with
an energy $e_0$ per particle and perturb it by increasing by some preassigned
$\Delta e$ the energy of a subset of adjacent particles. Let $e(x,t)$ be the
energy profile evolving from such a perturbed initial condition. 
We then ask how the perturbation
\be
f(x,t) = \langle e(x,t) - e_0 \rangle
\label{fdef}
\ee
behaves in time and space. The angular brackets denote an ensemble average 
over independent trajectories. Because of energy conservation,
$\sum_i f(i\ell,t) = \Delta e$  at all times, so that $f$ can be
interpreted as a probability density (provided that it is also
positive-defined and normalized). 

At sufficiently long times, one expects that $f(x,t)$ scales as
\begin{equation}
f(x,t)=t^{-\gamma}G(x/t^{\gamma}) \qquad \mbox{for $|x| < \tilde v t$}.
\label{eq:scail}
\end{equation}
for some finite velocity $\tilde v$. 

A sufficiently general framework where the above scaling behaviour is indeed
observed is that of Levy walks, where $f(x,t)$ is the probability density of
finding a particle whose kinetics is defined as follows. The particle moves
balistically in between successive ``collisions'' whose time
separation is distributed according to a power law, $\psi(t) \propto
t^{-\mu -1}$, $\mu >0$, while its velocity is chosen from a symmetric
distribution $\Psi(v)$. By assuming a $\delta$-like distribution
($\Psi(v) =(\delta(v-\tilde v) + \delta(v+\tilde v))/2$), the propagator
$P(x,t)$ (the probability distribution function to find in $x$ at time $t$, a
particle initially localized at $x=0$) can be written as
$P(x,t) = P_{L}(x,t) + t^{1-\mu}[\delta(x-\tilde vt) + \delta(x-\tilde vt)]$
where \cite{blumen}
\begin{equation}
  P_L(x,t) \propto
  \begin{cases}
    \begin{array}{ll}
         t^{-1/\mu}\, \exp \left[-(\eta x/t^{1/\mu})^2\right]
                                       & |x| \lesssim t^{1/\mu} \\
         t \, x^{-\mu-1}  & t^{1/\mu} \lesssim |x| <  \tilde v t\\
         0   &|x| > \tilde v t
     \end{array}
   \end{cases}
   \label{LW}
\end{equation}
Smoother, choices of the velocity distribution $\Psi(v)$ lead to different
broader side-peaks, but do not affect the shape and the scaling behaviour of
the bulk contribution $P_L(x,t)$ which scales as predicted in Eq.~(\ref{eq:scail})
with the exponent $\gamma=1/\mu$. Notice also that the two further parameters
$\tilde v$ and $\eta$ can be scaled out by using proper units of measure for the
time and space variables. Nevertheless, we shall see in the following that they
can carry some useful information when approaching a transition point.

From the evolution of the perturbation profile, it is possible to infer the
growth rate $\alpha$ of the thermal conductivity (see Eq.~(\ref{kappa})).
In fact, in Ref.~\cite{klaft}, it has been found that $\alpha$, the scaling
behaviour of the mean square displacement
($\sigma^2(t)= \sum_{i}i^{2}f(x=i\ell,t) \,\propto\, t^{\beta}$
and $\gamma$ are linked by the following relationships,
\footnote{
Although $\beta$ is here different from $1/k_BT$, we use this letter
for consistency with previous publications. We are anyhow confident that
it is always clear from the context which $\beta$ we are referring to.})

\be
\alpha\; =\; \beta\;-\;1\;=\; 2\;-\;\frac{1}{\gamma} 
\label{alfa2}.
\ee 
In particular, we see that the case $\gamma=1/2$ corresponds to normal
diffusion ($\beta=1$) and to a normal conductivity ($\alpha=0$). On the other
hand, $\gamma=1$ corresponds to a ballistic motion ($\beta=2$) and to a linear
divergence of the conductivity ($\alpha =1$).

\subsection{Infinitesimal perturbations}

With reference to 1D systems, it is convenient to introduce the
variable $\tilde u_i = \sqrt{m_i} u_i$ and express the perturbed trajectory as
$\tilde u_i^p= \tilde u_i^0+\delta_i$, where $\tilde u_i^0(t)$ is an equilibrium trajectory.
Since in gas-like systems there is only the kinetic contribution to the energy,
Eq.~(\ref{fdef}) can be written as
\begin{equation}
f(x=i\ell,t) = \langle (\tilde u_i^p)^2 \rangle - \langle (\tilde u_i^0)^2 \rangle  =
  \langle \delta_i^2 \rangle + 2 \langle \tilde u_i^0 \delta_i \rangle .
\label{rms}
\end{equation}
In the limit of infinitesimal perturbations, $\delta_i$ follows the tangent
space dynamics, i.e. the equation of motion linearized around a reference
trajectory. For the hard-point chain this is formally equivalent to that
in real space, since the dynamical equations are piecewise linear (the
transformations (\ref{coll}) describing the elastic collisions are indeed
linear). Accordingly, the energy conservation in real space transforms itself
into the conservation of the Euclidean norm
$Q=\sum_{i}f(i,t)=\sum_{i} \delta_{i}^{2}(t)$ of a generic perturbation.
This, in turn, implies that hard-point systems are not chaotic in the usual
sense of the word (all Lyapunov exponents are equal to 0).
Moreover, it is reasonable to conjecture (and we have verified numerically) 
that the dynamics in tangent space is uncorrelated with that in real space,
i.e. $\langle \tilde u_i^0 \delta_i \rangle = 0$. As a result, the energy
diffusion can be studied by simply looking at the spreading of infinitesimal
perturbations
\begin{equation}
f(x=i\ell,t) = \langle \delta_i^2 \rangle
\end{equation}
This approach has the advantage of dealing with positive-defined quantities
$\delta_i^2$ and thus allows getting rid of the strong fluctuations affecting
positive/negative variables. It is important to notice that this is possible
precisely because of the lack of standard deterministic chaos: in chaotic
systems, perturbations would exhibit an exponential growth, which can be
cancelled only after averaging over an unthinkable number of realizations.

\begin{figure}[!ht]
\begin{center}
\includegraphics[clip,width=8cm]{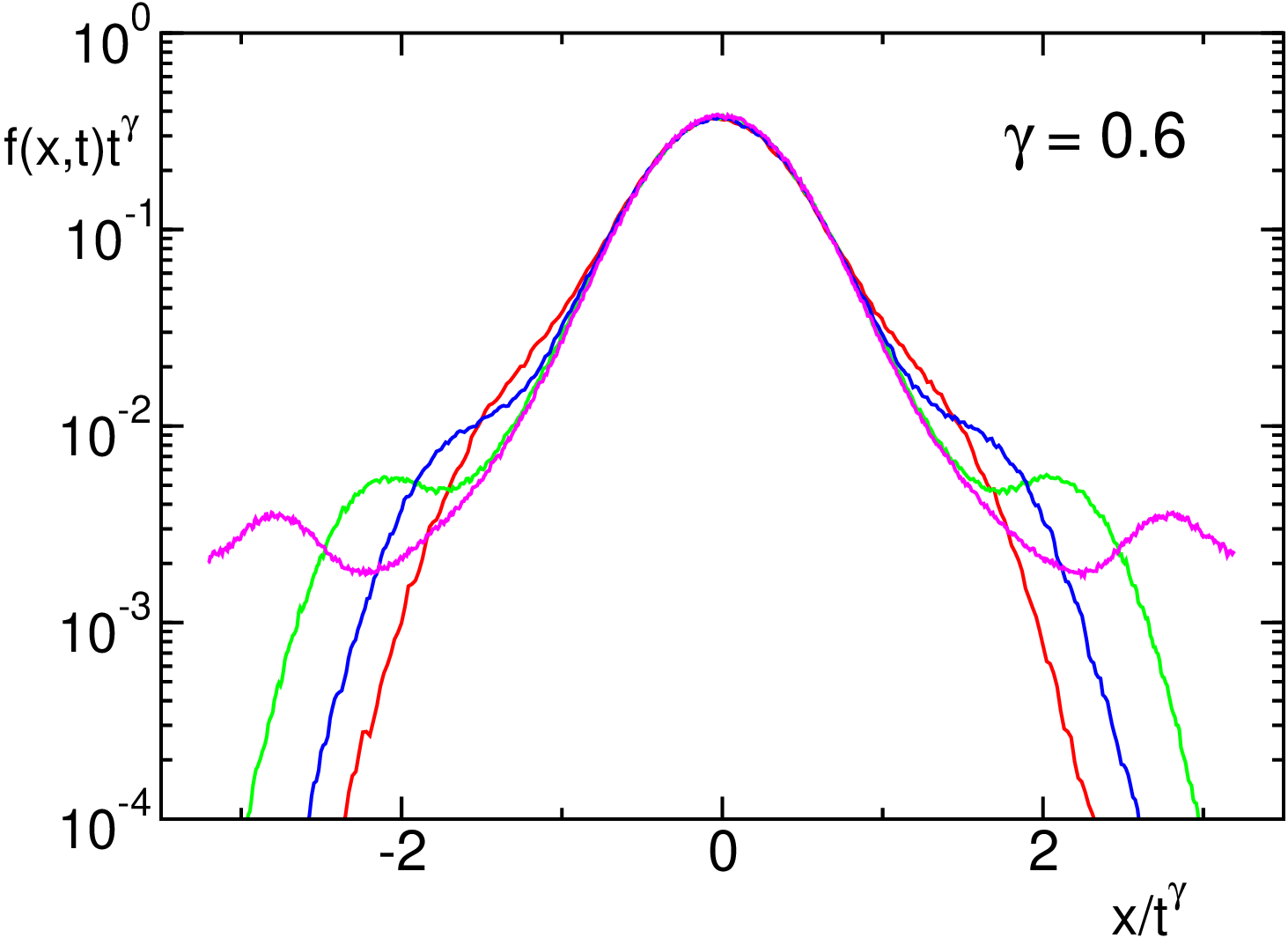}
\caption{Rescaled perturbation profiles, for the evolution in the
tangent space for $N=2045$ and $\gamma=0.6$, at $t=40$, 80, 160, 320 for
$\ell=0.2$. The profiles have been obtained by averaging over $10^4$
realizations.}
\label{profili0.2}
\end{center}
\end{figure}

We have first computed $f(x,t)$ for the hard--point chain with 
an average interparticle distance (specific volume) $\ell=0.2$, a value
sufficiently small to expect a perfect agreement with the behaviour of the HPG.
In fact, in Fig.~\ref{profili0.2}, we see that $\gamma=0.6$ leads to a
very good data collapse in the central region (the side peaks obviously
move away, as their position grows linearly in time).

These results confirm the behaviour of the HPG described in Ref.~\cite{denis},
where it was also found a perfect agreement with the scaling function
predicted for a Levy walk with $\mu =5/3$ (see Eq.~(\ref{LW})). In fact, 
the only
appreciable difference with the theoretical prediction, the broad secondary
peaks exhibited by $f(x,t)$, disappears as soon as the $\delta$-Dirac
distribution for ${\cal P}(v)$ is replaced by a Gaussian distribution with a
suitable width.

\begin{figure}[!ht]
\begin{center}
\includegraphics[clip,width=8cm]{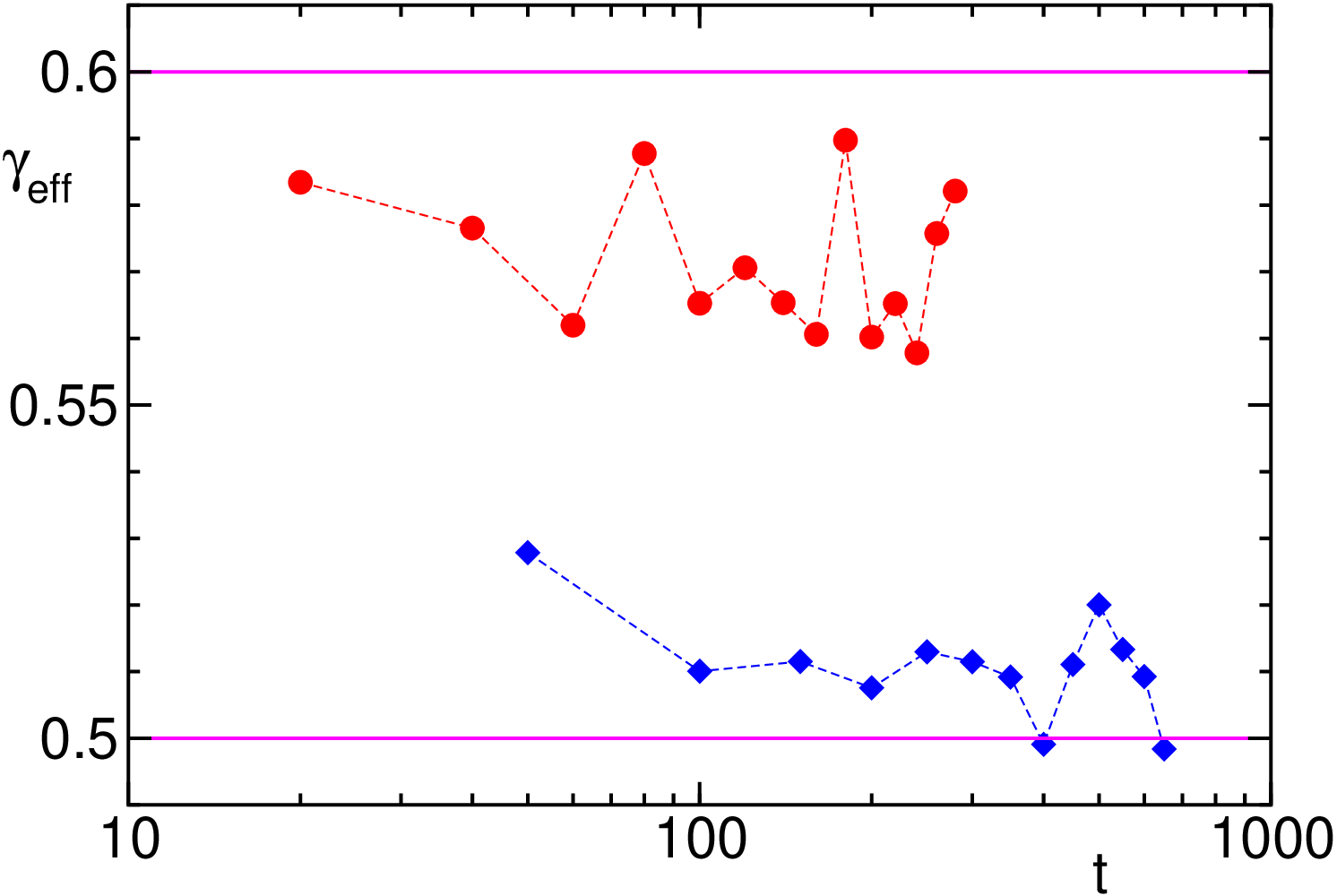}
\caption{The effective exponent $\gamma_{eff}$, Eq.(\ref{gamef}), 
versus time for two different
specific volumes: $\ell=0.3$ (circles) and $\ell=0.5$ (diamonds), $N=2045$}
\label{max_scal}
\end{center}
\end{figure}

Upon increasing the specific volume $\ell$, the effect of finite--distance
collisions (that are the novel ingredient of our model) becomes increasingly 
important. One may thus wonder whether the Levy walk scaling ~(\ref{LW}) 
still describes the perturbation dynamics. To answer this,    
in Fig.~\ref{max_scal}, we have plotted the logarithmic derivative
\be
   \gamma_{eff} =  -\frac{d \log f(0,t) }{d \log t}
\label{gamef}
\ee
of the maximum height for two values of the specific volume. For $\ell=0.3$
$\gamma_{eff}$ remains close to $0.6$, though slightly smaller.
As it is implausible that the asymptotic value of $\gamma$ depends smoothly
on any parameter, we attribute the deviations to finite-size corrections.
On the other hand, the deviations from 0.6 observed for $\ell =0.5$
are much stronger: the vicinity of $\gamma_{eff}$ to 1/2 is strongly suggestive
of a normal diffusion. 

\begin{figure}[!ht]
\begin{center}
\includegraphics[clip,width=8cm]{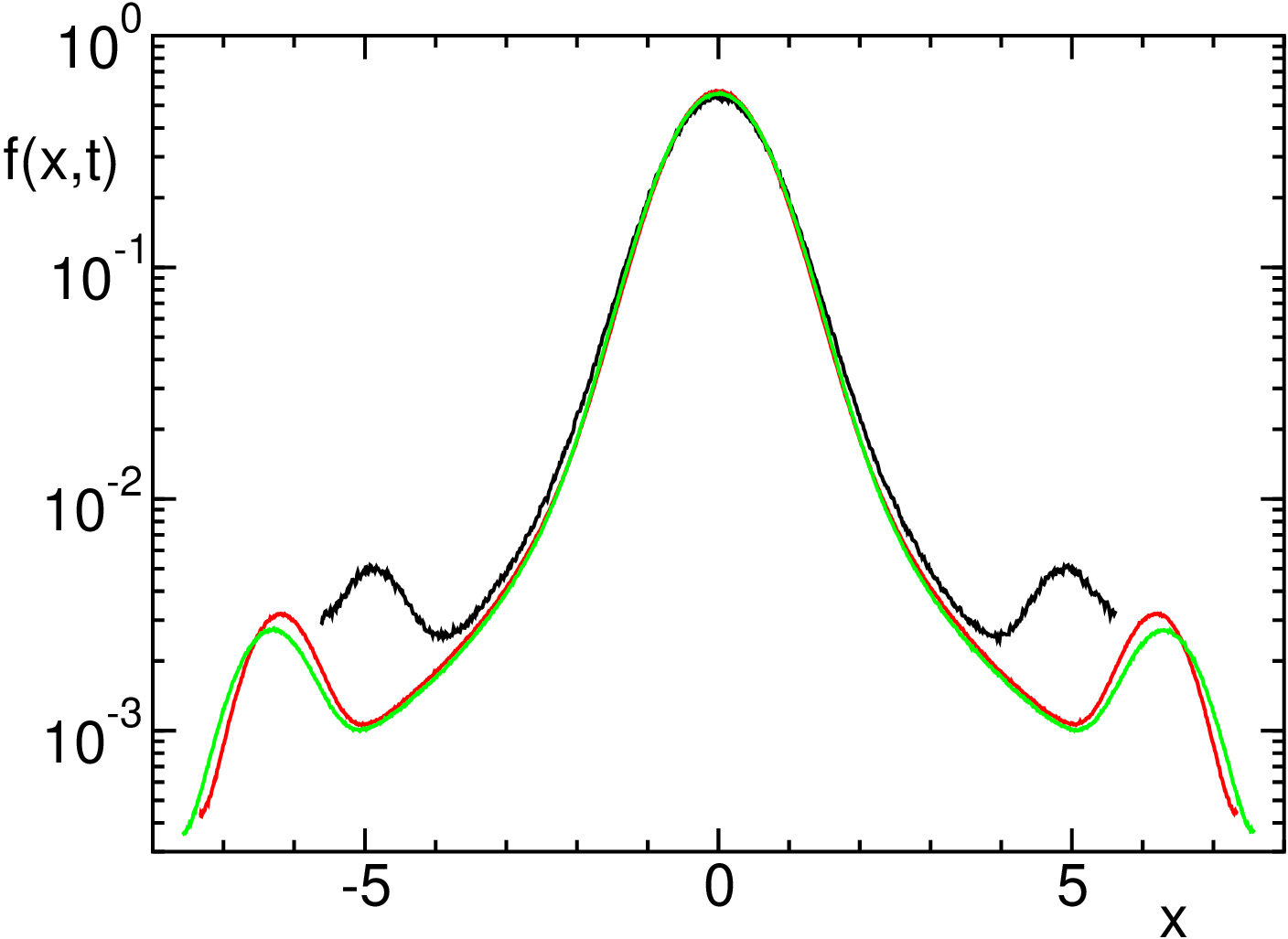}
\caption{The rescaled profiles for $\ell=0.2$ (at time $t=160$), $\ell = 0.3$
($t=280$), and $\ell=0.4$ ($t=580$). In each case $x$ and $f$ are scaled to
yield a unit variance and unit area.}
\label{prof_all}
\end{center}
\end{figure}

In order to further clarify the dependence on the specific volume, we have
plotted in Fig.~\ref{prof_all} the profiles corresponding to three different
specific volumes. The $x$ axis is scaled so as to yield a unit
variance. The good overlap among the various profiles confirms that all
cases where the pressure differs from zero exhibit the same universal
behavior (the different position of the side peaks reflects simply the
different timing and the different ballistic velocities in the three cases).

\begin{figure}[!ht]
\begin{center}
\includegraphics[clip,width=8cm]{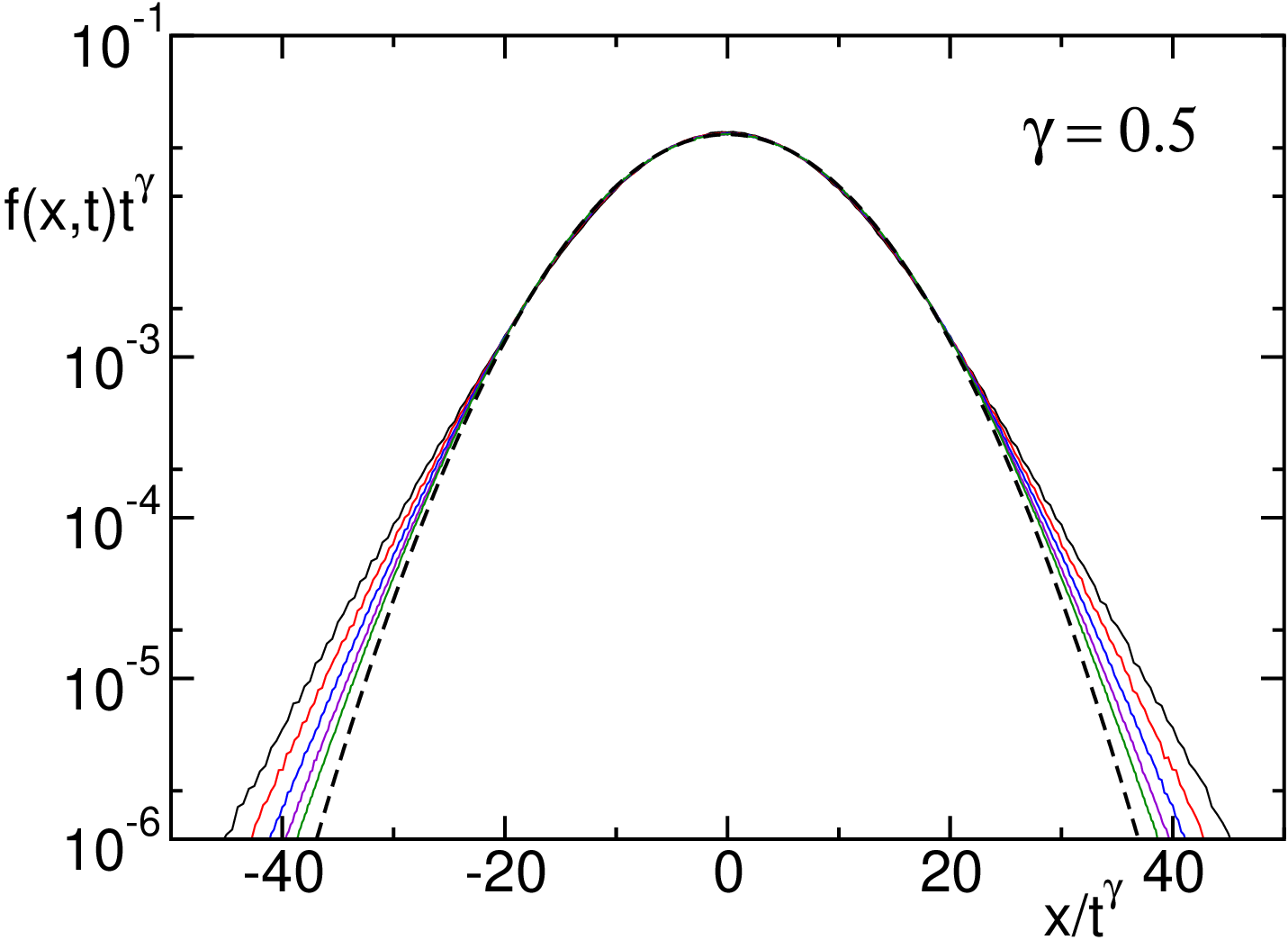}
\caption{Rescaled perturbation profiles for $\ell=0.5$, $N=2045$, at $t=50$,
100, 200, 400. and 750. The profiles have been obtained by averaging
over $10^4$ realizations. The dashed line corresponds to a Gaussian.}
\label{profili0.5}
\end{center}
\end{figure}

This is to be complemented by the observation that the perturbation profile
exhibits, in the zero-pressure limit a clearly different behaviour
(see Fig.~\ref{profili0.5} where the profiles are scaled according to
$\gamma=1/2$). No side peaks are present in this case and the bulk appears to
convergence towards a Gaussian profile (see the dashed curve, added for
reference).

Altogether, all these results suggest that the zero-pressure case is singular
and belongs to a different universality class as expected from the mode-coupling
theory. This is further confirmed by the behaviour of the parameter $\eta$
(see Eq.~(\ref{LW}))~. A convenient way to define $\eta$ is by scaling $x$ in such
a way that the variance of $f(x/t^\gamma)t^\gamma$ is equal to 1. As a result,
we find that $\eta(\ell)$ diverges, while approaching $\ell=1/2$. A fit of the
few available points is consistent with an inverse square root divergence, but
many factors prevent considering this prediction too seriously.

Although this result is a clear confirmation that the zero-pressure case
corresponds to a different universality class, it poses a problem, since
normal diffusion should correspond to normal conductivity. This prediction is
totally in contrast with both the numerical results for the FPU-$\beta$ system
and the solution of the mode-coupling equations which instead suggest a stronger
divergence for the conductivity.

\subsection{Finite perturbations}

In order to understand the puzzling behavior of infinitesimal perturbations,
it is worth exploring the dynamics of finite rather than infinitesimal
perturbations. In fact, the signature of ``stable chaos", a type of irregular
behaviour found in linearly stable coupled map lattices, is precisely the
different propagation rate for finite and infinitesimal perturbations \cite{PT}.
Moreover, since it was already argued in Ref.~\cite{denis} that the HPG might be
a Hamiltonian model exhibiting stable chaos, it is not unreasonable to expect a
different behaviour for the two classes of perturbations.

With reference to Eq.~(\ref{rms}), we follow pairs of trajectories
$(\tilde u_i^0(t),\tilde u_i^p(t))$, where the first element is a typical
unperturbed equilibrium configuration and the second element is obtained from
the first one by adding a given amount of energy in the three sites around the
origin. We then study the evolution of the perturbation by computing the
ensemble average of the instantaneous local distance
$\Delta(x,t) = \langle (\tilde u_i^p(t) - \tilde u_i^0(t)^2\rangle $
and the ensemble average of the energy difference
$\Delta e(x,t) = \langle (\tilde u_i^p(t))^2\rangle - \langle (\tilde
u_i^0(t))^2\rangle$.

\begin{figure}[!ht]
\begin{center}
\includegraphics[clip,width=8cm]{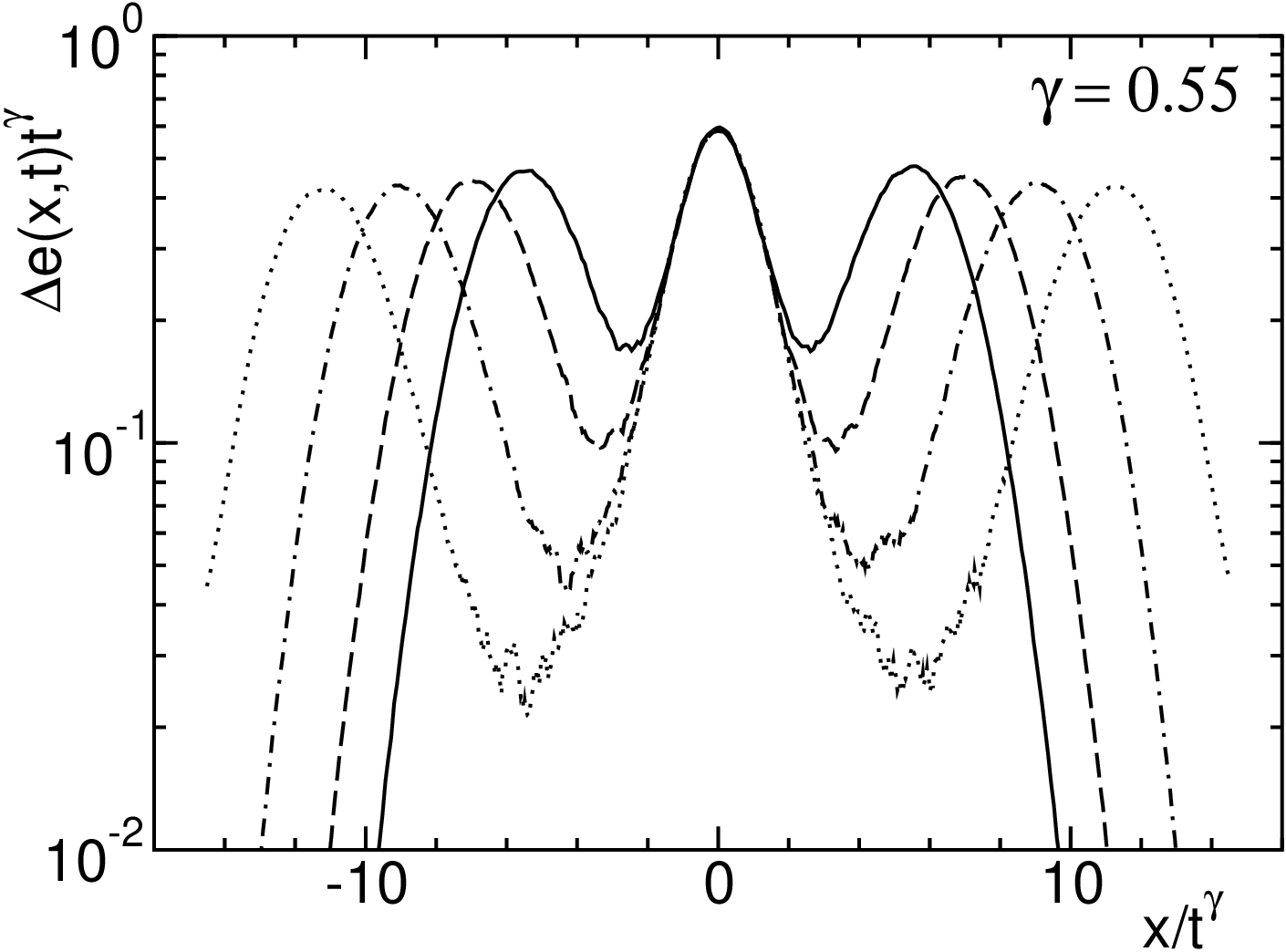}
\caption{Energy perturbation profiles for an energy density $e=1$, at 
$t=5$, 10, 20, and 35 ($N=2045$), for $\ell=0.2$. The profiles
have been obtained by averaging over $10^4$ realizations.}
\label{prof_fin_02}
\end{center}
\end{figure}

From the results reported in Fig.~\ref{prof_fin_02}, the central peaks show
a good overlap for $\gamma = 0.55$. Considering the shorter time scales
with respect to those ones achieved in the corresponding tangent-space
simulations, the deviation from $0.6$ cannot be taken as an indication
of a different scaling behaviour. A substantial difference is instead found
by comparing the velocity of the side peaks which is definitely larger
in this second context (see the next section for a detailed discussion).
The too short time scales prevent a quantitative comparison with the predictions
of the Levy walk model: in fact the central peak should decay by another
order of magnitude to appreciate deviations from the Gaussian behaviour.

\begin{figure}[!ht]
\begin{center}
\includegraphics[clip,width=8cm]{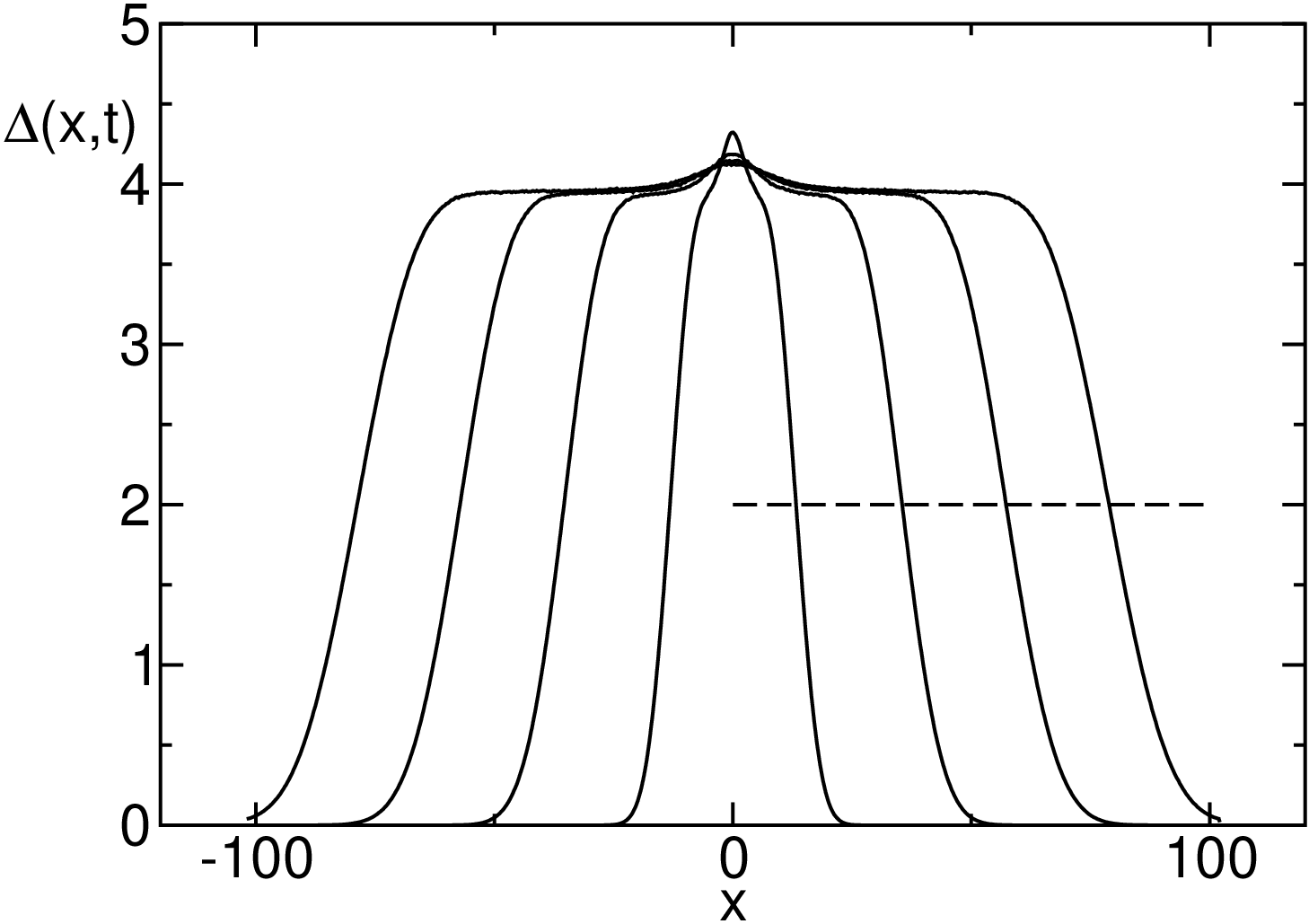}
\caption{Propagation of finite amplitude perturbations for $N=2045$, 
energy density $e=1$ and $\ell=0.2$. The profiles have been obtained 
by averaging over $10^4$ realizations. Different curves are at times
$t=5$, 10, 20 and 35 respectively.}
\label{pert_fin}
\end{center}
\end{figure}

In Fig.~\ref{pert_fin} we plot the amplitude of the perturbation
itself. The propagation velocity is estimated by determining the half-height
position of the front (see the horizontal dashed line) at different times.
The velocity $c_{fa}$ turns out to coincide with that of the side peaks in
the set-up of Fig.~\ref{prof_fin_02}.

\begin{figure}[!ht]
\begin{center}
\includegraphics[clip,width=8cm]{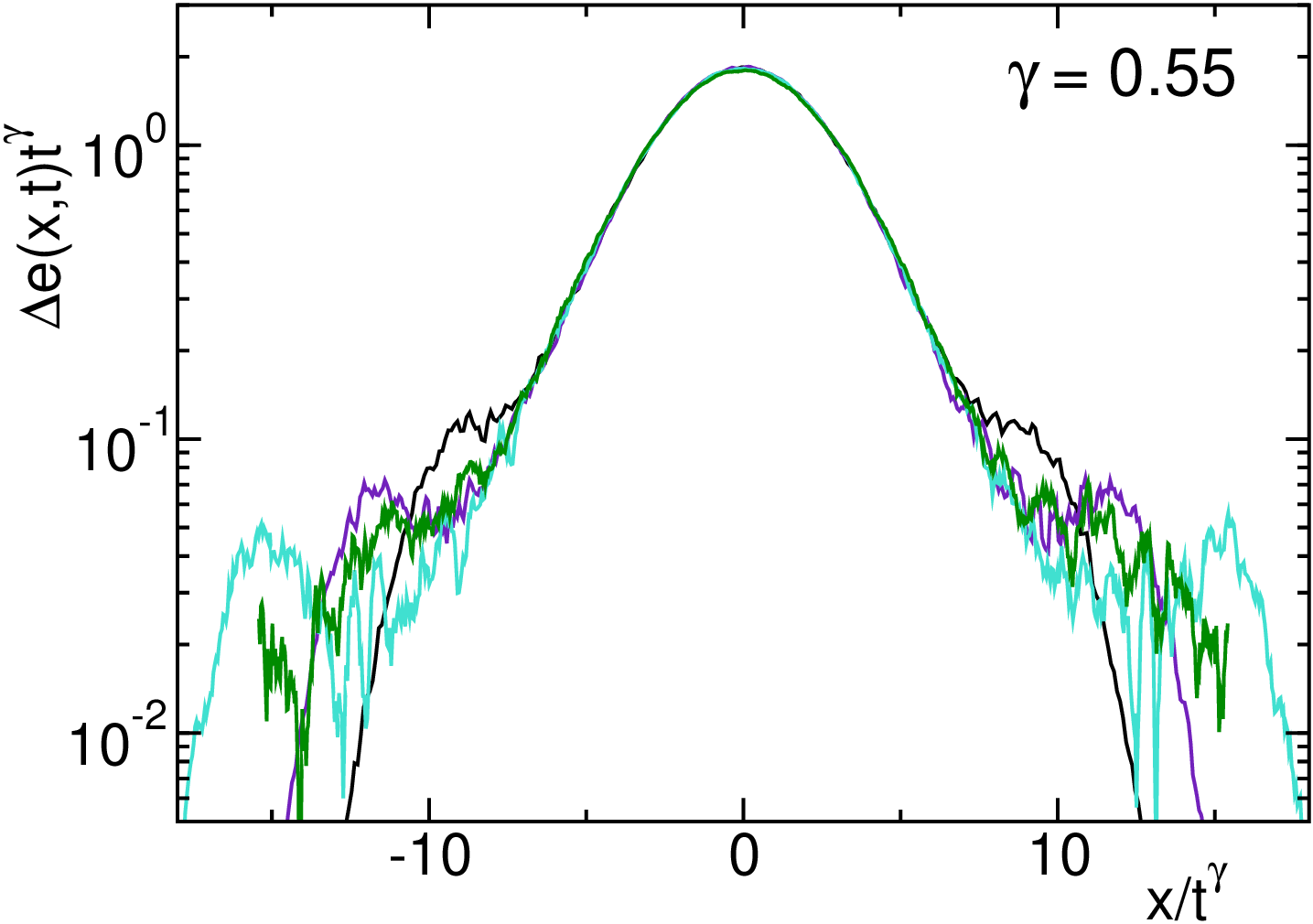}
\caption{Rescaled energy perturbation profiles for an energy density $e=1$, 
at $t$=12, 24, 48, 96 ($N=2045$), for $\ell=0.5$, $\gamma$=0.55. The profiles
have been obtained by averaging over $10^4$ realizations.}
\label{prof_fin_05}
\end{center}
\end{figure}

Finally, we have studied the case $\ell = 0.5$. As it can be seen in
Fig.~\ref{prof_fin_05}, in the available time range we do not appreciate
any difference with respect to the previous non-zero pressure case.
Moreover, in agreement with tangent space results, the side peaks are
basically absent. Altogether, fluctuations in real space are so strong
that it is almost impossible to extract any useful information, apart from
the velocity of the perturbations themselves, that is the only observable
unambiguously different with respect to that of infinitesimal perturbations.

\section{Ballistic phenomena}

In the previous section we have encountered two different velocities for
the propagation of infinitesimal and finite perturbations.
From a thermodynamic point of view, there exist two further velocities: 
the adiabatic $c_{ad}$ and the isothermal $c_{is}$ sound velocity, which 
are defined respectively as
\be
c_{ad}^2= \Big ( \frac{\partial P}{\partial \rho} \Big )_{S} \qquad \qquad
c_{is}^2= \Big ( \frac{\partial P}{\partial \rho} \Big )_{T}
\label{c2}
\ee
where $\rho$ is the mass density. They are related by
\be
c_{ad}^2\;=\; \frac{C_P}{C_V} \; c_{is}^2
\ee
where $C_{P}$ and $C_{V}$ are the constant-pressure and constant-volume
specific heats.

\begin{figure}[!ht]
\begin{center}
\includegraphics[clip,width=8cm]{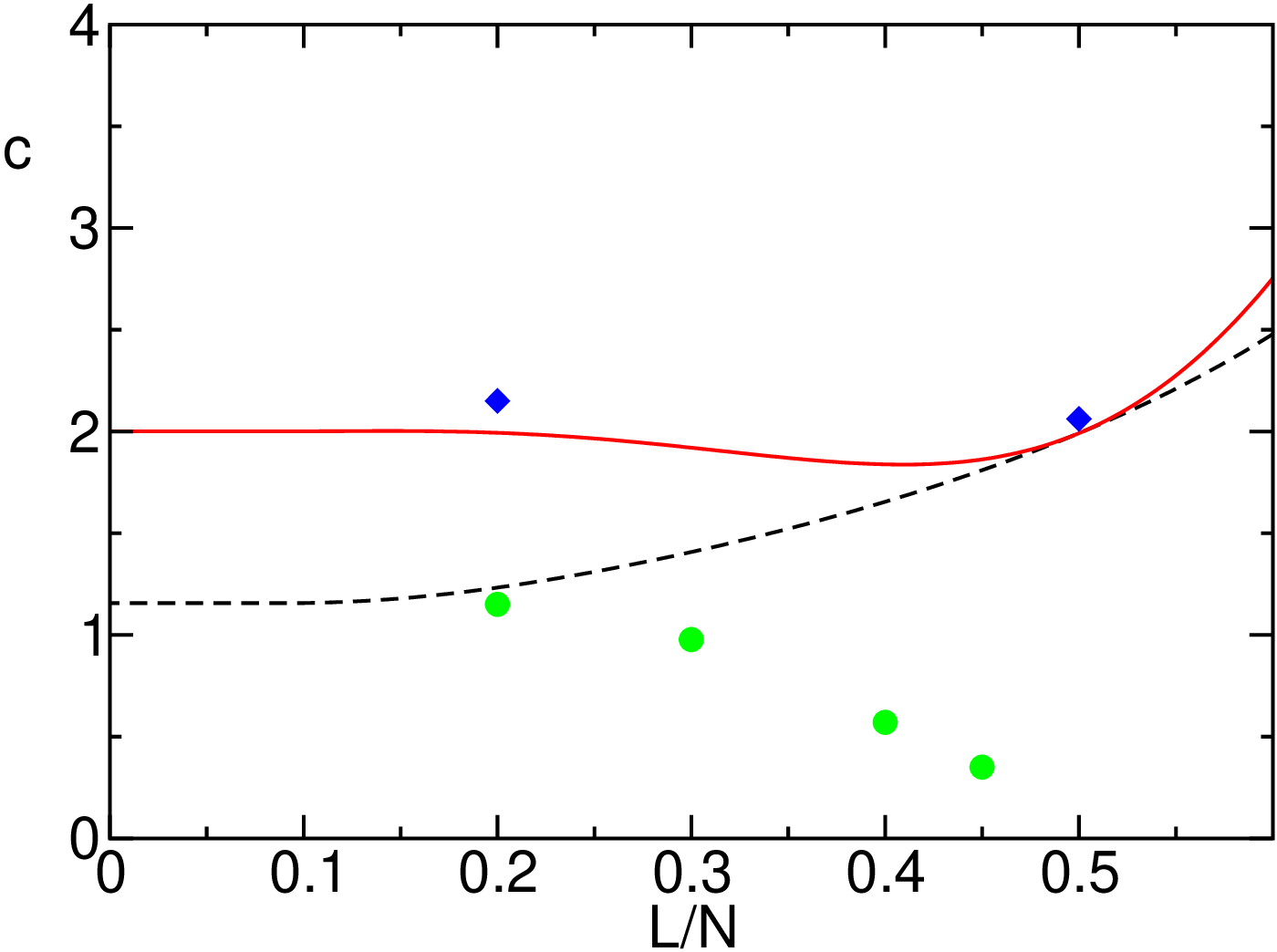}
\caption{Thermodynamic velocities versus the specific volume $\ell = L/N$ for the
hard point chain. Solid and dashes curves correspond to the adiabatic and
isothermal velocity, respectively. The solid circles correspond to the
propagation velocity as determined from the behaviour of infinitesimal
perturbations, while the diamonds correspond to the velocity of finite
amplitude perturbations.}
\label{veloc}
\end{center}
\end{figure}

From the equation of state (\ref{eq:state}), we can determine both velocities 
for the hard-point chain. They are reported in Fig.~\ref{veloc} versus the 
specific volume. For $\ell \to 0$, the hard-point chain reduces to the 
HPG model. In this limit, a simple expression can be obtained for the two 
velocities. The equation of state reduces to
\be
PL\;=\;k_{B}NT\;=\;\frac{2\rho}{1+r}\frac{k_{B}T}{m}
\ee
where $\rho=m(1+r)L/2N$. From Eq.~(\ref{c2}) we obtain,
\be
c_{is}^2=\frac{2k_{B}T}{(1+r)m} \qquad \qquad
c_{ad}^2=3\;c_{is}^2
\ee
because $C_P/C_V=3$ in the HPG. By inserting the parameters used in the
the numerical simulations we obtain $c_{is}=1.154$ and $c_{ad}=2$
for $r=2$, and $c_{is}=1$, $c_{ad}=1.722$ for $r=3$.

We can now compare all such velocities. From the position of the side peaks for
different specific volumes, we obtain the tangent--space velocity $c_{ta}$
(circles in Fig.~\ref{veloc}). In the HPG limit, $c_{ta}$ coincides with
$c_{is}$ (we have checked that this holds true also for different mass ratios).
However, upon increasing the specific volume $c_{is}$ increases, while $c_{ta}$
decreases and vanishes when $\ell$ approaches 0.5. Besides confirming the
peculiarity of the zero-pressure case, these simulations reveal that $c_{ta}$
differs from both thermodynamical velocities.

Furthermore, we have determined the velocity $c_{fa}$ of finite-amplitude
perturbations (diamonds in Fig.~\ref{veloc}), finding that it is close to,
though slightly larger than $c_{ad}$~. The difference is 
larger than the statistical fluctuations, but we cannot exclude that
this is due to systematic deviations that vanish at longer times. We have
verified that $c_{fa}$ agrees also with the velocity estimated from the
dispersion of long-wavelength Fourier modes.\footnote{The agreement claimed
in Ref.~\cite{denis} between the tangent space and adiabatic velocity was the
result of an error in the normalization.}

\begin{figure}[!ht]
\begin{center}
\includegraphics[clip,width=8cm]{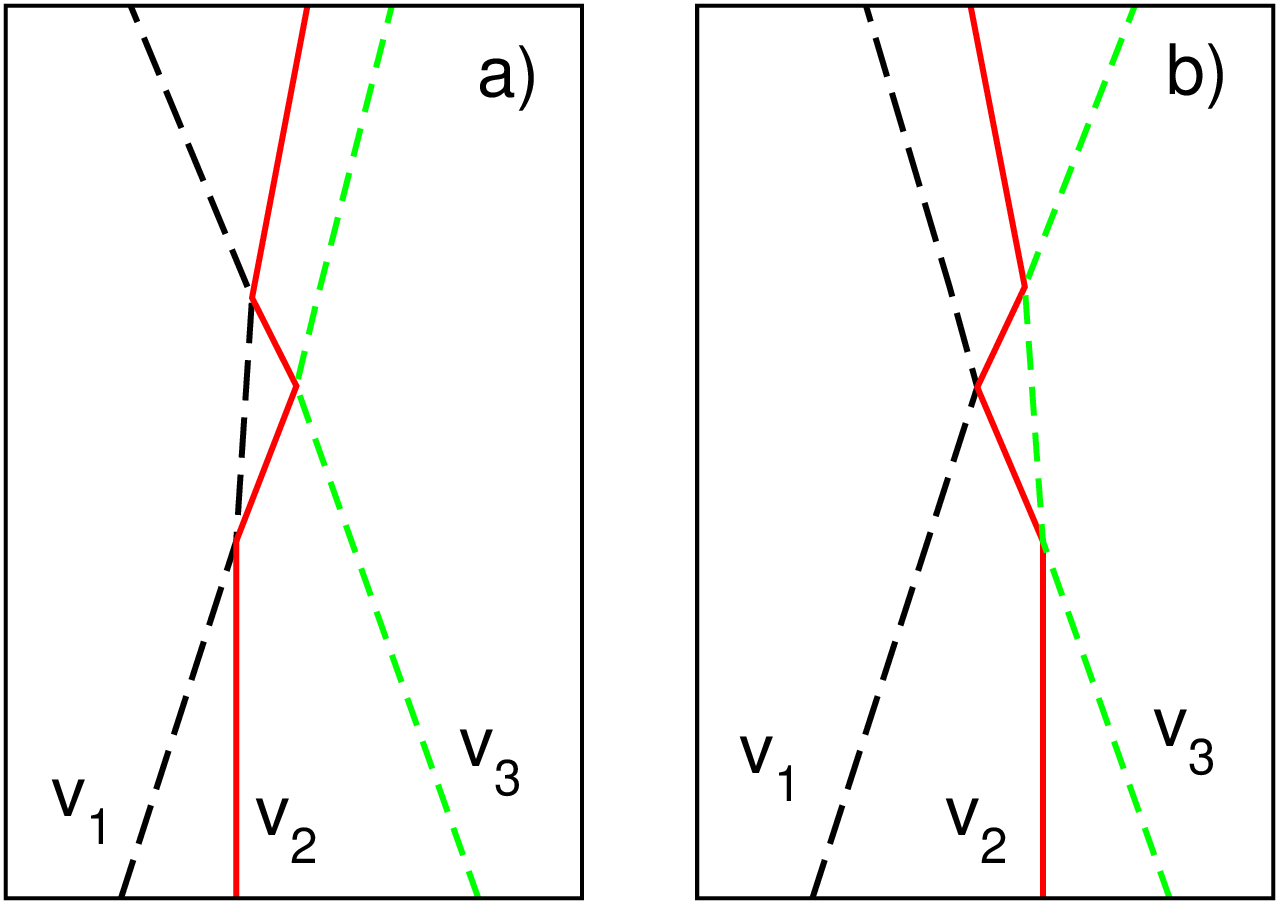}
\caption{The evolution of three neighbouring particles in the vicinity of
a three body collisions.}
\label{scatt}
\end{center}
\end{figure}

From the dynamical point of view,  the most important observation is however
that $c_{fa}$ is definitely larger than $c_{ta}$. This inequality is the typical
signature of ``stable chaos". In order to strenghten this conclusion we show in
Fig.~\ref{scatt} another typical feature of stable chaos, i.e. that
self-sustained irregular behaviour is accompanied by the presence of
discontinuities in phase-space dynamics. By discontinuiy, we mean a codimension
one manifold across which an infinitesimal difference in the initial conditions
is amplified to a finite value in a finite time \footnote{Actually, as shown in 
Ref.~\cite{BP}, the condition is not so strict and strong localized 
nonlinearities may suffice.}. In Fig.~\ref{scatt}, we show that this indeed
occurs in the HPG in the vicinity of a three--body collision. By shifting the
central particle to the right, one discontinuously passes from the sequence of
collisions depicted in panel {\it a}  to the sequence depicted in panel {\it b}.
Since the effects of the collisions do not commute, one has a discontinuous
change in the outcoming velocities, while passing through the configuration
which yields the three body collision. In the vicinity of such discontinuities,
small but finite perturbations are suddenly amplified and this provides a
mechanism for sustaining irregular behavior in the absence of an exponential
amplification of infinitesimal differences.

\section{The energy current}

In view of the difficulties encountered in extracting the scaling behavior
of thermal conductivity from direct energy diffusion simulation, we now
turn our attention to the fluctuations of the total heat flux.
Our past experience with other 1D models in fact suggests that this observable
is much more reliable since the fluctuations due to the spatial dependence
are integrated out.

In order to properly treat hard-point chains, it is necessary to implement the
general microscopic definition of the heat flux \cite{LLP03}
\be j_{i} =
\frac{1}{2}(x_{i+1}-x_{i})(u_{i+1}+u_{i})F(x_{i+1}-x_{i})
+ \frac{1}{2} m_i \dot{x}_{i}^3  \equiv j_i^{(1)} + j_i^{(2)} 
\label{flusso}
\ee
where $F$ denotes the force exerted by the $i$--th particle on its right
neighbor. In the HPG, only the second term matter, since interactions occur
only when the relative distance between the neighbouring particles
$x_{i+1}-x_{i}$ vanishes. On the contrary, in the hard-point chain, 
``collisions'' occur also at the distance $a$. However, since the force 
is singular, one cannot implement the definition (\ref{flusso}) as it stands. 
By defining the
force between two particles as the momentum difference induced by a collision,
$j_i^{(1)}$ can be written as the kinetic energy variation times the actual
distance $a$, i.e., $j_i^{(1)} = a m_i(u_{i}^{'2}-u_{i}^2)/2$ divided by a
suitable time-interval $\Delta t$. In order to get rid of the microscopic
fluctuations, it is necessary to consider a sufficiently long $\Delta t$, so
as to include a large number of collisions. Since the number of collisions is
proportional to the system size, it is only in long systems that fluctuations
can be removed without spoiling the slow dynamics of the heat flux. By fixing
$\Delta t = 1$ in our arbitrary units, it is sufficient to simulate a few 
hundreds of particles.

\begin{figure}[!ht]
\begin{center}
\includegraphics[clip,width=8cm]{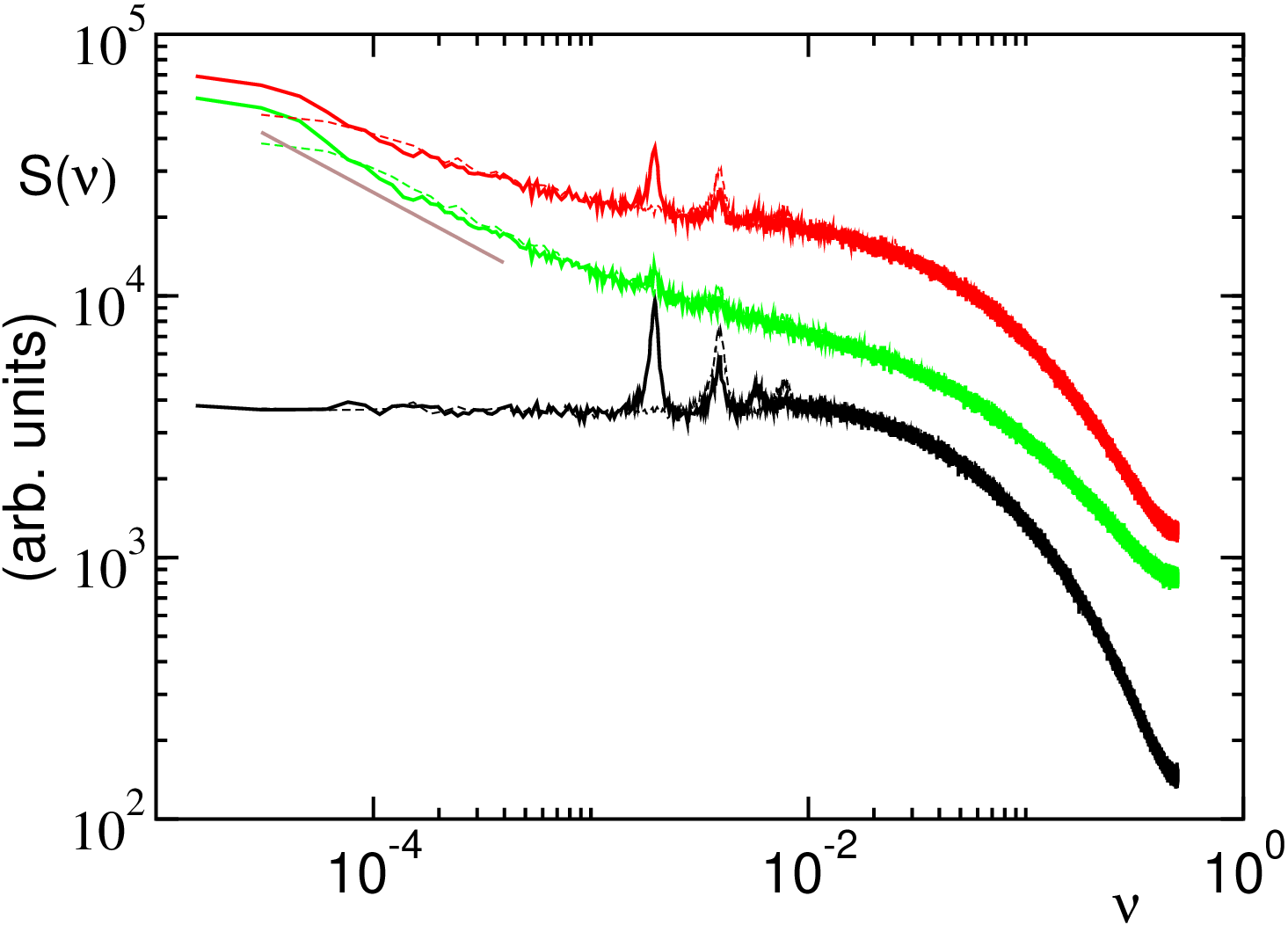}
\caption{Power spectrum of the heat flux for energy density $e=1$ and specific volume $\ell=0.5$. 
Solid and dashed lines refer to $N=2045$ and $N=1021$ respectively. From bottom to top: the contribution
due to $j^{(2)}$, $j^{(1)}$ (see Eq.~(\ref{flusso}), and their sum. Power spectrum have been obtained 
by averaging over $10^4$ realizations.}
\label{spettri}
\end{center}
\end{figure}

Besides computing the power spectrum $S(\nu)$ of the total flux, we have
determined the power spectrum of the two separate contributions $j^{(1)}$ and
$j^{(2)}$ for $\ell = 0.5$. The $j^{(2)}$ contribution, which exhibits
$\nu^{-1/3}$  power-law divergence in the HPG, here appears to saturate (see the
lowest solid curve in Fig.~\ref{spettri}). From the comparison with the spectrum
obtained for a shorter chain length (see the dashed curve), it is clear that the
saturation is not a finite-size effect. Accordingly, one might be tempted to
conclude that the zero-pressure case is characterized by a normal conductivity
as also suggested by the simulation performed in tangent space. However, the
spectrum of the other contribution $j^{(1)}$, shows that this is not such
conclusion is incorrect. In fact, the middle solid curve exhibits a
low-frequency divergence which is terminated by a finite-size saturation (as
revelead by looking at the spectra corresponding to two different chain
lengths). A power-law fit within the central frequency range  (i.e. away from
the saturation region) yields an exponent around 0.45 (see the  straigh segment,
shifted for the sake of clarity). This value is not only definetely larger than
the 1/3 value predicted for nonzero pressure, but it is also fairly close to the
results found for the FPU-$\beta$ model \cite{LLP03b}.  Although we cannot claim
a quantitative confirmation of the 1/2 prediction of mode-coupling theory, the
conjecture that zero-pressure systems do belong to a different universality
class is further strengthened.

On the other hand, one should not forget that the heat flux definition 
requires summing the two contributions we have separately analysed.
By doing so, not unexpectedly, the resulting power spectrum exhibits
an intermediate behaviour (see the top curve in Fig.~\ref{spettri}).
It is reasonable to conjecture that the resulting slower convergence is
a finite-size correction due to the strong constant background due to
the $j^{(2)}$ contribution, but it is also honest to admit that these
results altogether confirm the difficulty of extracting reliable estimates
in zero-pressure systems.

\section{Adding internal degrees of freedom}

\begin{figure}[!ht]
\begin{center}
\includegraphics[clip,width=8cm]{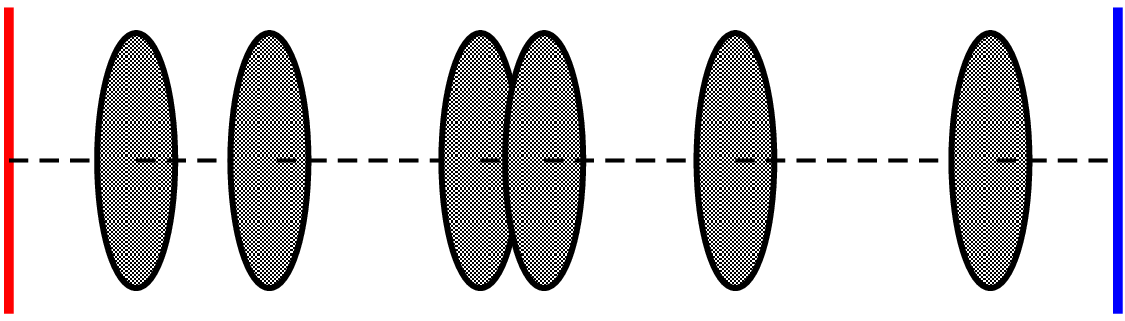}
\caption{Sketch of the hard disk model}
\label{hard_disk}
\end{center}
\end{figure}

For the sake of completeness, we have explored another variant of the
HPG that is obtained by introducing an additional degree of freedom to each
particle in the original model. As it is sketched in Fig.~\ref{hard_disk},
particles are replaced by infinitesimally thin disks, which can rotate along
the $x$ axis and thereby possess an angular velocity as well as an angular
momentum that we denote with $\zeta_i$. Like in the original HPG model,
collisions do occur whenever the longitudinal position of two neighbouring
particles coincides and the dynamics is determined by the corresponding
collision rules. Besides linear momentum $P_i$ and energy $E_i$, there is
another conserved quantity, namely angular momentum $Z_i$. For the sake of
simplicity, we consider an equal-disk gas, while the units of measure are
chosen in such a way that both the mass and the momentum of inertia are set
equal to 1. Altogether, the quantities
\begin{eqnarray}
P_i &=& p_{i}+p_{i+1} \nonumber \\
Z_i &=& \zeta_{i}+\zeta_{i+1}\\ 
E_i &=& p^2_{i}+\zeta^2_{i}+p^2_{i+1}+\zeta^2_{i+1}
\end{eqnarray}
are left invariant by the collision. One can easily verify that the following
transformations satisfy the above constraints,
\begin{eqnarray}
p'_{i} &=& p_{i}\sin^2 \phi + p_{i+1} \cos^2 \phi + (\zeta_{i+1}-\zeta_i) \sin
\phi \cos\phi   \nonumber \\
p'_{i+1} &=& p_{i}\cos^2 \phi + p_{i+1} \sin^2 \phi - (\zeta_{i+1}-\zeta_i) \sin
\phi \cos \phi  \label{sto} \\
\zeta'_{i} &=& \zeta_{i}\cos^2 \phi + \zeta_{i+1} \sin^2 \phi + (p_{i+1}-p_i) \sin
\phi \cos \phi  \nonumber \\
\zeta'_{i+1} &=& \zeta_{i}\sin^2 \phi + \zeta_{i+1} \cos^2 \phi - (p_{i+1}-p_i) \sin
\phi \cos \phi  \nonumber
\end{eqnarray}
where $\phi$ is a free parameter. One might determine a deterministic updating
rule for $\phi$ by assuming a specific interaction Hamiltonian during the
collision. For the sake of simplicity, we have preferred to randomly extract
$\phi$ according to a flat distribution restricted to the angles which
guarantee that the outgoing velocities change direction (to avoid particle
crossing). 

Although this model has been developed by assuming that $\zeta_i$ is an angular
momentum, it equally applies to the case where $\zeta_i$ is a linear momentum,
like for a chain embedded in a planar geometry with transverse
periodic boundary conditions. Actually, this latter interpretation would make
the model almost identical to that one studied in \cite{DN03}.

Like before, we have studied the evolution of an energy packet in tangent
space, $f(x,t)= \delta p^{2}_{i}(t)+\delta \zeta^{2}_{i}(t)$, having the care
of referring to the same sequence of $\phi$ values in the real as well
as in tangent space.

\begin{figure}[t]
\begin{center}
\includegraphics[clip,width=8cm]{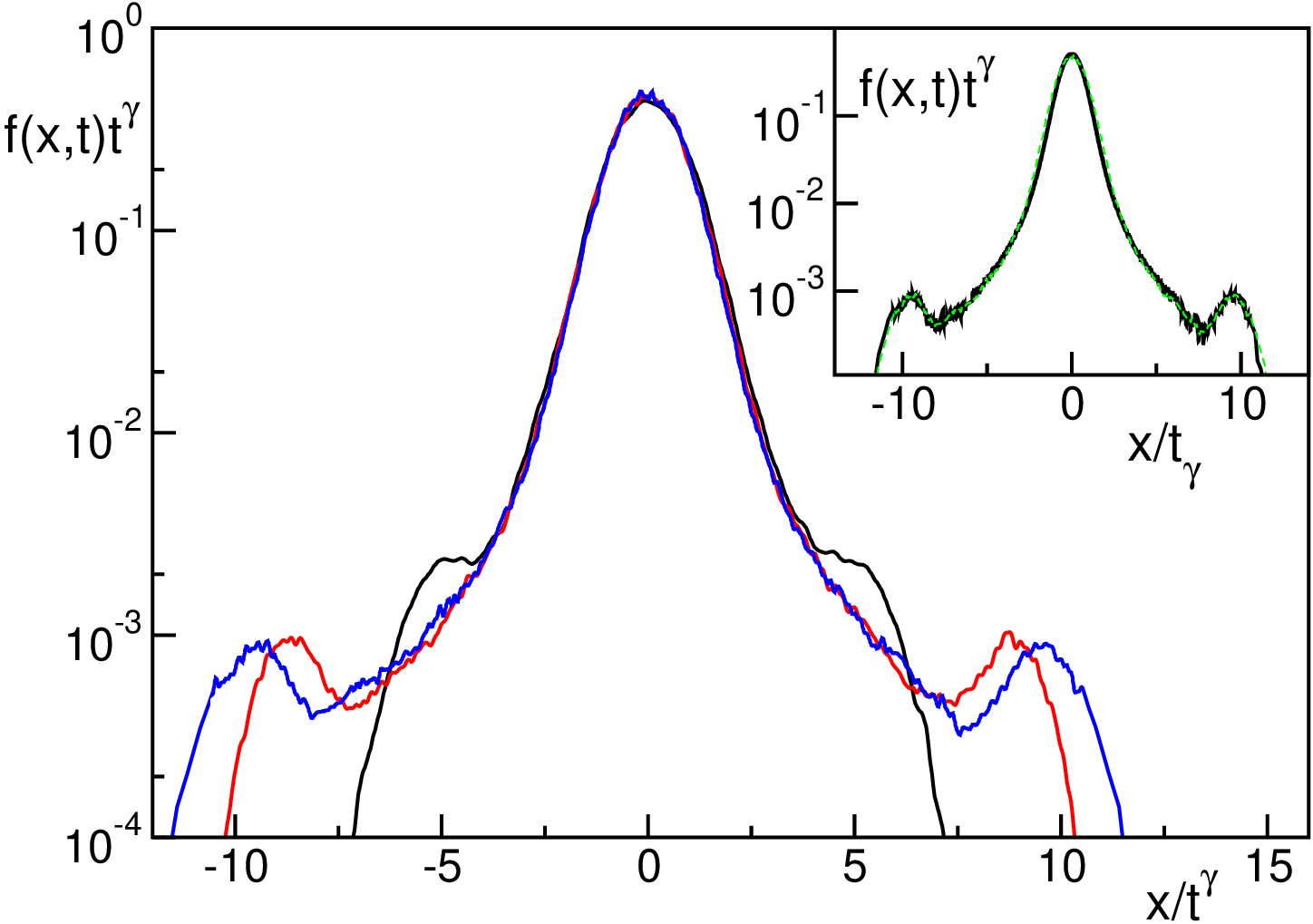}
\caption{ Rescaled perturbation profiles, for the evolution in the
tangent space, at $t= 80, 160, 240$ for the generalized HPG,
Eqs.(\ref{sto}). The profiles have been obtained by averaging over
$10^4$ realizations. In the inset, the profile at $t=160$ (solid
line) is compared with the propagators of a Levy walk for an
exponent $\mu=5/3$ with a Gaussian distribution with average
equal to 1 and rms 0.0347)(dashed line).} 
\label{genHPG}
\end{center}
\end{figure}

The quantitative result of numerical studies are very similar to the
diatomic HPG \cite{denis} and to the hard-point chain model with $\ell=0.2$.
More precisely, the evolution in the tangent space is in a
good agreement with the Levy--walk model with the same scaling
exponent $\gamma=0.6$ (Fig.~\ref{genHPG}). 

Numerical value of the velocity propagation is consistent with the isothermal
sound velocity of the hard-point chain model with $\ell=0.2$ and equal masses.

\section{Conclusions and open problems}

In this paper we have seen that the hard-point chain model allows extending the
same approaches developed for the hard-point gas to a system, where the
pressure can be tuned from positive to negative values. We have restricted
our analysis to nonnegative pressure values, because the evolution for negative
pressures is symmetrically related to that for positive values (see, e.g., the
equation of state as plotted in Fig.~\ref{eqstato}).
Although, the various numerical techniques have confirmed the difficulties
in extracting reliable estimates of scaling rate, we have been able to
convincingly show that the zero-pressure case belongs to a different
universality class. The most compelling evidence is arising from the tangent
space evolution: (i) the scaling exponent $\gamma$ is close to 1/2 instead of
3/5 as for the hard point gas; (ii) the velocity $c_{ta}$ goes to zero.

Another general observation follows from the different propagation velocity
exhibited by finite and infinitesimal perturbations. We interpret this fact as
an indication that the behavior of hard-point chains is an instance of ``stable
chaos", i.e. an irregular dynamical regime sustained by discontinuities in
phase-space rather than by the exponential amplification.

Besides these two general comments, all conclusions about the scaling
behaviour require further checks and a more careful analysis (if doable at all).
For instance, although it is quite reasonable to assume that the scaling
rate $\gamma$ is always equal to $3/5$ whenever the pressure is different
from 0, it should be verified that the systematic deviations observed for
the specific volume $\ell = 0.3$ decay when going to longer times.

\begin{figure}[t]
\begin{center}
\includegraphics[clip,width=8cm]{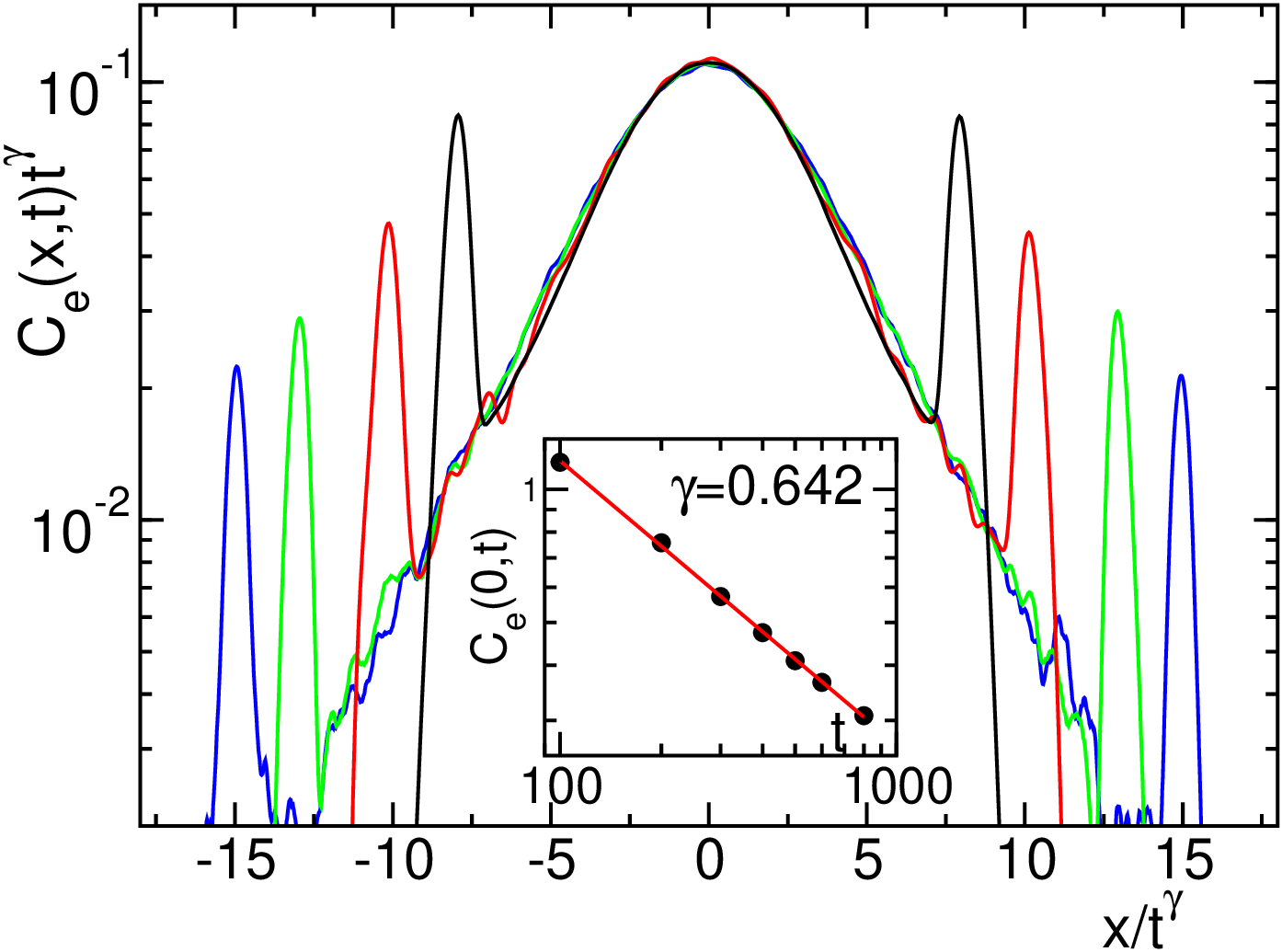}
\caption{Rescaled energy correlation function for the FPU-$\beta$ model, for
an energy density $e=1$. The different profiles, which correspond to times
$t= 100$, 200, 400, and 600, have been obtained by averaging over
$10^4$ realizations. In the inset the decay of the maximum is reported versus
time together with a best power-law fit.}
\label{prof_fpu}
\end{center}
\end{figure}

On the other hand, the difficulties encountered while performing simulations
with finite amplitude perturbations indicate that there is little room to make
progress, by adopting this approach. In this sense, it is perhaps a better
idea looking directly at the behaviour of the correlation function of
the energy density \cite{zhao},
\begin{equation}
C_e(x,t) = \langle \delta e(y,\tau)\delta e (x+y,t+\tau)\rangle \quad ,
\end{equation}
where the angular brackets denote a spatial as well temporal average.
At $t=0$, the correlation function $C_e(x,0)$ is a $\delta$ function in
space. Moreover, in the microcanonical ensemble, energy conservation implies 
that the area $\int dx C_e(x,t)$ is constant at all times. By therefore 
assuming that $C_e(x,t)$ is normalizet to have unit area, its behaviour is 
formally equivalent to that a
diffusing probability distribution. This observation has led Zhao 
to determine the scaling behaviour of the heat conductivity from the
growth rate of the variance of $C_e(x,t)$ \cite{zhao}. As the determination
of the variance is troubled by the fluctuating tails, it is preferrable to
proceed like in the previous sections, by looking at the decay of the
maximum $C_e(0,t)$ that is statistically more reliable. 
For the sake of providing a concrete example of such an approach, we 
performed a  microcanonical simulations of an FPU-$\beta$ chain in a 
lattice of 2048 particles.  The correlations $C_e(x,t)$ for different times, 
are reported in Fig.~\ref{prof_fpu}.
They have been rescaled by choosing $\gamma=0.642$, a value that has been
determined from a best fit of the maximum of the correlation function.
From the Levy-walk machinery, it follows that $\alpha = 2- 1/\gamma \approx
0.44$. This value is significantly different from 1/3 and definitely
confirms that zero-pressure systems do belong to a different universality
class. On the other hand $0.44$ is still not so close to the mode-coupling
prediction (1/2) and thus confirms once more the presence of strong finite-size
corrections which certainly need be understood before making conclusive
claims.

\section*{Acknowledgments}

This work is supported by the PRIN2005 project {\it Transport properties of
classical and quantum systems} funded by MIUR-Italy. Part of the numerical
calculation were performed at CINECA supercomputing facility through the project
entitled {\it Transport and fluctuations in low-dimensional systems}.

\end{document}